\documentclass[journal]{IEEEtran}
\usepackage{amsmath,amssymb,amsthm,lipsum} %调用数学模式宏包%
\usepackage{cite}
\usepackage{algorithmic}
\usepackage{algorithm}
\usepackage{graphicx}
\usepackage{textcomp}
\usepackage{stfloats}
\usepackage{booktabs}
\usepackage{bm}
\usepackage{subfigure}
\usepackage{setspace}
\usepackage{color}
\usepackage[T1]{fontenc}
\usepackage{cases}
\allowdisplaybreaks[4]
\ifCLASSINFOpdf
\else
\fi
\theoremstyle{Theorem}

\theoremstyle{remark}

\theoremstyle{Definition}

\theoremstyle{Lemma}

\theoremstyle{Corollary}

\theoremstyle{Proposition}

\begin{document}


\title{Near-Field Communications with 
	Block-Dominant Compressed Sensing: 
	Fundamentals, Approaches, and Future Directions}

\author{Liyang~Lu,~\IEEEmembership{Member,~IEEE,}
	  Ke Ma,
	  Yue Wang,~\IEEEmembership{Senior Member,~IEEE,}
	      Zhaocheng~Wang,~\IEEEmembership{Fellow,~IEEE}\\	         %
\thanks{
	This work was supported in part by the National Natural Science Foundation
	of China under Grant 62471275, in part by the National Science 
	Foundation under Grant 2413622, in part by the Postdoctoral Fellowship
	Program of China Postdoctoral Science Foudation (CPSF) under Grant
	GZC20231374, and in part by the China Postdoctoral Science Foundation
	under Grant 2024M751680. (\textit{Corresponding author: Zhaocheng Wang.})
	
	L.~Lu, K.~Ma and Z.~Wang are with Tsinghua University, Beijing 100084, China (e-mails: luliyang@mail.tsinghua.edu.cn, ma-~k19@mails.tsinghua.edu.cn, zcwang@tsinghua.edu.cn).
	
	Y. Wang is with the Computer Science Department, Georgia State University, Atlanta, GA 30303, USA (e-mail: ywang182@gsu.edu).

} %
\vspace*{-5mm}
}


\markboth{}%
{Shell \MakeLowercase{\textit{et al.}}: Bare Demo of IEEEtran.cls for IEEE Journals}



\maketitle

\begin{abstract}
In the context of extremely large-scale antenna arrays deployed in sixth-generation (6G) mobile networks, near-field (NF) communications have gained considerable attention. Unlike the planar waves formulated in the far-field, electromagnetic radiation propagates as spherical waves in the NF. This alteration affects the NF channel characteristics, particularly resulting in weak sparsity in angular-domain NF channels, which poses tricky challenges to the application of compressed sensing (CS). Motivated by these facts, the block-dominant compressed sensing (BD-CS) techniques are proposed to assist NF communications. This article starts with the introduction on why block sparsity exists in the distance-limited NF region. Then, block-dominant side-information (BD-SI) is exploited to facilitate the actual NF communication implementation.
While BD-CS shows promise in providing exceptional channel estimation accuracy and high spectral efficiency, several key challenges, opportunities, and practical implementation issues in NF communications need careful consideration.
\end{abstract}

\IEEEpeerreviewmaketitle

\section{Introduction}\label{introduction}

The sixth generation mobile networks are expected to support enhanced capacity and wide coverage by relying on extremely large-scale antenna arrays \cite{dai2023cm,yuanbin2023tvt}, as exemplified by holographic and extremely large-scale multiple-input multiple-output communications \cite{wangzhe2024}. As a result, the so-called Rayleigh distance \cite{dai2022tcom} which represents the coverage-limit of the near-field (NF) is substantially increased, possibly to hundreds of meters \cite{zhu2023cl}. Explicitly, it is proportional to the square of the array aperture and the carrier frequency. Within the Rayleigh distance, NF propagation is dominant, where electromagnetic radiation has to be modeled by spherical waves.

Due to the intrinsic distance ingredient, the phase of each antenna element in the steering vector turns to be a non-linear function of the antenna index, potentially causing  energy spread effect in NF channel \cite{dai2022tcom}. Hence, the number of non-zero channel-taps becomes high in the angular domain. Moreover, the scatterers are likely to be aggregated in the NF region \cite{FundamentalsofWirelessCommunication}. The combined energy spread and scatterer cluster effect results in a high proportion of non-zero channel-taps, i.e., the channel impulse response becomes weakly sparse. When the channel impulse response is sparse, as in the conventional mmWave systems, compressed sensing (CS) is usually preferred both in channel estimation (CE) and transmit precoding (TPC). However, for weakly sparse channel impulse responses, CS cannot be readily applied.

	Fortunately, the NF channel can be transformed into a relatively sparse embodiment by involving the polar-domain transform (PDT) proposed in \cite{dai2022tcom}. Especially, the discrete Fourier transform (DFT) matrix for angular-domain mapping consists of FF steering vectors, while the NF steering vectors form the PDT matrix \cite{wuhaochen2024}. However, the PDT mapping tends to significantly increase the number of the total channel-taps by sampling both the angle and distance ingredients, leading to an unaffordable complexity increase in both CE and TPC \cite{dai2022tcom}.
\textcolor{black}{As shown in \cite{dai2022tcom} and \cite{FundamentalsofWirelessCommunication}, the energy spread and scatterer cluster effects result in adjacent non-zero angle/distance values, which is regarded as a block structure \cite{liyang2022tsp}. The block CS (BCS) can effectively enhance sparse recovery performance by exploiting this structural characteristic.}

	In actual wireless channels, valuable information that exhibits the property of block structure, referred to as block-dominant side-information (BD-SI), can be utilized to enhance the recovery performance of BCS \cite{wuhaochen2024}. For instance, the non-zero channel-tap positions on different subcarriers are usually the same, exhibiting joint correlation property \cite{liyang2023arxiv}. BCS technique assisted by BD-SI is studied and regarded as block-dominant CS (BD-CS).
 
\begin{figure*}
	\centering
	\includegraphics[scale=0.58]{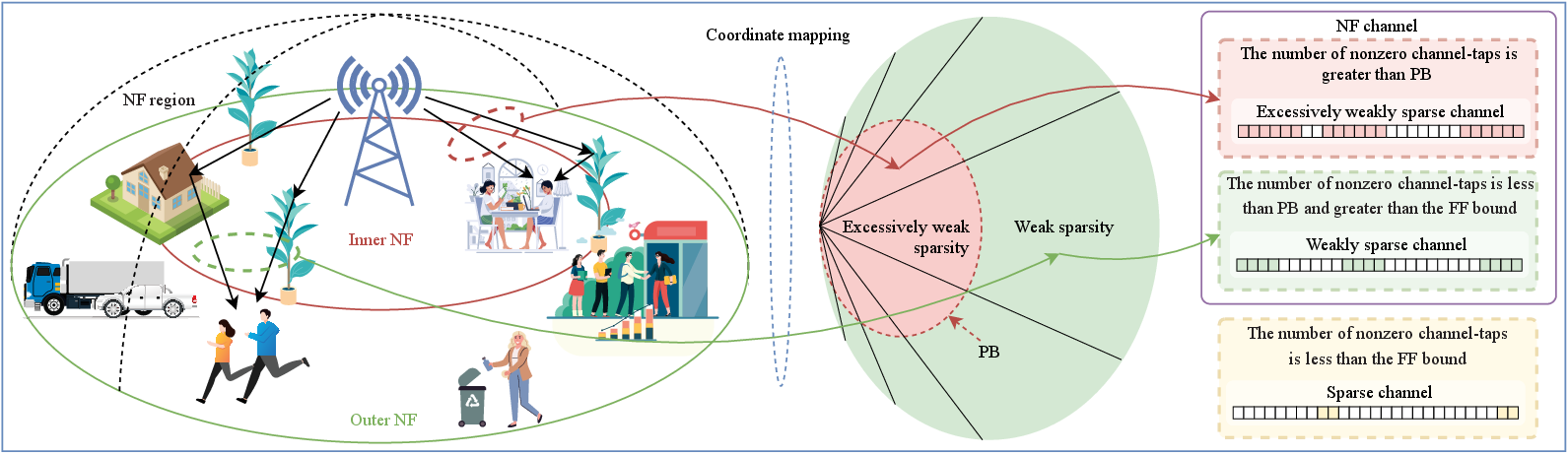}\\
	\caption{NF region partition.}\label{NF region}
\end{figure*}

Furthermore, the NF region is partitioned into inner and outer NF regions by an elaborate partition boundary (PB), as illustrated in Fig.~\ref{NF region}. The outer NF channel is only weakly sparse, hence potentially allowing BD-CS to be used in the angular domain. Meanwhile, the inner NF channel, which has a higher number of non-zero channel-taps, is excessively weakly sparse, where the conventional angular-domain mapping may not work. Thus, the PDT based BD-CS may be harnessed for the inner NF channel. 
However, the joint design of BD-CS and multi-domain alternating mapping, i.e., angular or polar domain mappings, for NF communications has numerous challenges that need to be handled carefully.

Given the aforementioned background and considering the corresponding challenges, we introduce an innovative BD-CS assisted NF communication concept, and our main contributions are summarized as follows.
\begin{enumerate}
	\item The block structure of NF channel is briefly explained. We elucidate both energy spread effect and scatterer cluster effect in NF communications, both of which contribute to the channel's block structure property. Weak sparsity and excessively weak sparsity are introduced to illustrate the fundamental characteristics of the NF channel. 
	
	\item BD-SI is introduced from three different perspectives, including consecutive weights from relevant channels, joint correlation from different subcarriers, and decaying nature from multi-path energy, which could provide the valuable information in NF communications.
	
	\item BD-CS is introduced to facilitate BD-SI aided NF communications, where NF region is partitioned into inner and outer regions, each exhibiting distinct sparsity characteristics. Both CE and hybrid TPC schemes utilizing BD-CS are proposed to improve the estimation accuracy and the spectral efficiency, respectively.
	
	\item Potential future research directions for NF communications with BD-CS are investigated, and some research outlooks are discussed accordingly, where the prospective insights relying on NF communication theory are provided, paving the way for practical solutions.
\end{enumerate} 

\section{BD-CS Fundamentals}\label{sec2}

\textcolor{black}{This section introduces NF channel properties, defines BD-CS, and then illustrates its key characteristics in NF communications. 		 }		

\subsection{\textcolor{black}{NF Channel Properties}}
\vspace{-3mm}
\textcolor{black}{\subsubsection{Energy spread effect}
For a spherical wave modeled NF channel, since the phase of each element in the NF steering vector is non-linear to the antenna index, several FF steering vectors are jointly utilized to describe one NF steering vector, causing energy spread effect \cite{dai2022tcom}. }

\textcolor{black}{\subsubsection{Scatterer cluster effect} As NF region is more limited than FF region, scatters and reflectors are not located at all directions from transmitter or receiver but are more likely to be grouped into several clusters than those in the FF \cite{FundamentalsofWirelessCommunication}.}

\textcolor{black}{In what follows, we describe weak sparsity and block structure of the NF channel stemming from these two effects. }

\subsubsection{Sparsity}  A channel exhibits sparsity when the number
	of non-zero elements is significantly smaller than its overall
	length \cite{yuanbin2023tvt}.

\textcolor{black}{\subsubsection{Weak sparsity and excessively weak sparsity}
	The number of the non-zero channel-taps of the NF channel in angular domain, where the angular-domain sparse basis DFT consists of FF steering vectors, is significantly increased due to the energy spread effect. Note that the closer to the BS, the more severe the energy spread effect. The NF channel exhibits weak sparsity when it is far from the BS, while it exhibits excessively weak sparsity, surpassing weak sparsity, when it is close to the BS. In both weakly sparse and excessively weakly sparse scenarios, conventional CS cannot provide satisfactory performance due to the high number of non-zero channel-taps.} 

	For a channel characterized by sparsity, weak sparsity, or excessively weak sparsity, if the number of non-zero elements in the channel is equal to $K$, we collectively refer to it as a channel with $K$-sparsity, and it is considered to be $K$-sparse.

\textcolor{black}{\subsubsection{Block structure and block sparsity} For the NF, block structure occurs naturally because of the following two reasons. Firstly, due to the energy spread effect, the energy of NF path spreads into multiple angles when a channel is transformed into an angular-domain representation, which is regarded as the block structure.  
	Secondly, stemming from the scatterer cluster effect, each cluster can contribute to a continuum of paths. A near-continuous path delay results in near-continuous angle of arrivals, constituting a block structure of the channel in angular domain. Similarly, the steering vector in the polar-domain sparse basis cannot exactly correspond to one NF steering vector, constituting the block structure. If the number of non-zero blocks in a channel is small enough, the channel exhibits block sparsity. Since considering the block sparsity of a channel can enhance the estimation performance with more reliability, it helps alleviate the issues of weak sparsity and excessively weak sparsity. }

\subsection{What is BD-CS?}
\begin{figure}
	\centering
	\includegraphics[scale=0.55]{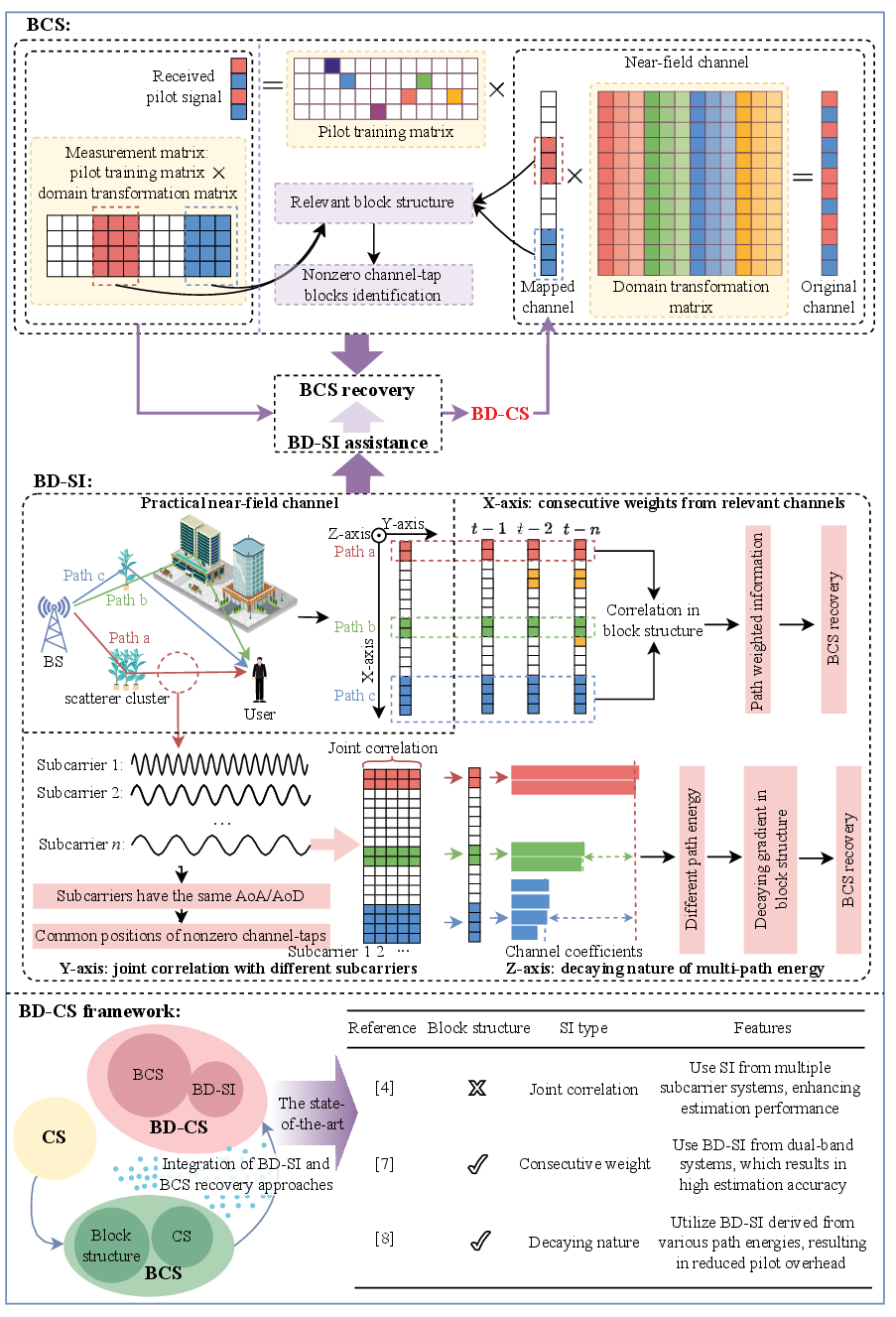}\\
	\caption{Illustration of BD-CS.}\label{3dBD-SI}
\end{figure}
In this subsection, CS and BCS techniques are first introduced, followed by the concept of BD-CS.

\subsubsection{CS technique} 
	\textcolor{black}{An essential prerequisite of adopting CS to NF communications is that the NF channel is sufficiently sparse.} By uniformly sampling the angular ingredient and non-uniformly sampling the distance ingredient, PDT achieves a sparser mapping of NF channel than its angular-domain counterpart. Hence, the NF channel can be acquired with much fewer samples than the overall dimension of the polar-domain representation. However, compared to DFT mapping, the dimension of PDT mapping is significantly enlarged, leading to high computational complexity of both CE and TPC when the conventional CS is adopted \cite{dai2022tcom}.

\subsubsection{BCS technique} 
\textcolor{black}{As illustrated in Fig. 2, CS turns into a more universal BCS technique, which can achieve higher accuracy for estimating channels exhibiting a block structure than that using conventional CS \cite{liyang2023arxiv}.} Explicitly, BCS retrieves continuous entries of the underlying signal rather than one single entry per iteration. \textcolor{black}{In this context, BCS boils down to CS, when setting the block length parameter in BCS as 1.}

\subsubsection{BD-CS technique} The framework of BD-CS is given in Fig. \ref{3dBD-SI}. It is a new BCS concept considering the block structure as the dominant characteristic, which is further assisted by the BD-SI, for signal recovery. 
\textcolor{black}{Side-information (SI) refers to a type of information, such as the underlying position and power information of non-zero elements in a channel, stemming from surrounding environments or communication systems. BD-SI is an extensive version of SI, which considers the intrinsic block structure of the signal to be analyzed.}
Explicitly, the BD-SI of a practical NF channel is presented in Fig. \ref{3dBD-SI} from three different perspectives, which is explained as follows.

\emph{Consecutive weights from relevant channels (x-axis)}:
In practice, the positions of the non-zero channel-tap blocks across consecutive time slots or in dual-band systems exhibit similarity \cite{Ali2018twc,wuhaochen2024}, resulting in a correlation of channel characteristics between these time slots or systems. This non-zero position information can be utilized to enhance CE and TPC within the current time slot or frequency band. Specifically, it can be mapped into  consecutive weight information based on certain function indicators, particularly the non-linear exponential function indicator used in \cite{liyang2023tcom}. Thus, the CE and TPC is reformulated as an effective weighted block-sparse recovery problem \cite{wuhaochen2024}.

\color{black}{\emph{Joint correlation with different subcarriers (y-axis)}:}  \textcolor{black}{For a multi-carrier communication scenario, the channels on different subcarriers can be used to generate pilot matrix simultaneously.} \color{black}{Meanwhile, the steering vectors of different subcarriers are usually the same, and thus the channels to be recovered on different subcarriers share a common block sparsity after the sparse domain mapping \cite{dai2022tcom}.} \textcolor{black}{This is referred as the block multiple measurement vector model, and CE and TPC are transformed into a block multiple measurement vector recovery problem correspondingly \cite{liyang2023arxiv}. The non-zero block information among channels on different subcarriers provide useful joint correlation, resulting in efficient CE and TPC.}

\textcolor{black}{\emph{Decaying nature of multi-path energy (z-axis)}:} The channel-taps of different paths exhibit amplitude differences, which is associated with a decaying gradient. \textcolor{black}{ This decaying gradient information can be acquired from previous time slot, as well as from dual-band systems that exhibit similar channel characteristics \cite{Ali2018twc}. By analyzing the potential decaying gradient of NF channel before transmission, the pilot overhead is significantly reduced by more assured reliability on identifying the correct positions and energy of the non-zero channel-tap blocks \cite{liyang2023arxiv,liyang2023tcom}.}

\textcolor{black}{Compared with BCS, the exploitation of BD-SI in BD-CS will provide improvements in both performance and pilot overhead, since BD-CS is more capable of identifying correct non-zero channel-tap blocks and beam directions. Furthermore, BD-CS can be applied not only in the NF but also in the FF, which is regarded as a universal technology that can even be applied to the scenarios where both NF and FF exist.}

\begin{table*}
	\centering
	\begin{tabular}{|p{3.95cm}|p{4.1cm}|p{4.1cm}|p{4.1cm}|p{4.1cm}}
		\multicolumn{4}{c}{TABLE I: Comparisons of various CE methods}\\
		\cline{1-4}    
		Methods & Key features & Advantages & Disadvantages \\ \cline{1-4}    
		(1) Angular-domain CS     & Estimate the FF channel which is sparse in angular domain & Perform well with low pilot overhead & Does not apply to the NF CE \\ \cline{1-4}
		(2) Polar-domain CS     & Estimate the NF channel after PDT & Higher accuracy than (1)  & High mapping complexity; severe weak sparsity when the distance ingredient is relatively large  \\\cline{1-4}
		\textcolor{black}{(3) Angular-domain BCS}     & \textcolor{black}{Estimate the block-sparse FF channel, and the NF channel with mild energy spread effect} & \textcolor{black}{Medium complexity and accuracy among angular-domain methods}  & \textcolor{black}{May not work when the energy spread effect is severe} \\ \cline{1-4}
		\textcolor{black}{(4) Polar-domain BCS} & \textcolor{black}{Estimate the block-sparse NF channel after PDT }  & \textcolor{black}{Medium complexity and accuracy among polar-domain methods}  & \textcolor{black}{Higher complexity than (3) due to PDT}\\ \cline{1-4}
		\textcolor{black}{(5) Angular-domain BD-CS}     & \textcolor{black}{Exploit the BD-CS to estimate the NF channel which is weakly-sparse}  & \textcolor{black}{The best accuracy when the distance ingredient is relatively large; the lowest complexity among all the methods compared}     & \textcolor{black}{Additional calculation complexity of  BD-SI}      \\ \cline{1-4}
		\textcolor{black}{(6) Polar-domain BD-CS}     & \textcolor{black}{Use BD-CS and PDT to estimate the over weakly-sparse channel} & \textcolor{black}{The best accuracy when the distance ingredient is relatively small; the lowest complexity among the polar-domain methods}  & \textcolor{black}{Additional calculation complexity of BD-SI; higher complexity than (5) due to PDT}\\ \cline{1-4}
	\end{tabular}\label{tab1}
\end{table*}

\subsection{Why BD-CS?}\label{II-C}

The essential reason for inaccurate estimation is that the number of non-zero channel-taps is larger than the upper bound of sparsity required for reliable recovery \cite{liyang2022tsp}. BD-CS has two key features to address this issue by significantly improving this upper limit, namely the joint use of the block structure and the BD-SI. 
As proved in several works, e.g., \cite{liyang2023arxiv} and \cite{liyang2022tsp}, when the number of non-zero channel-taps is fixed, the block structure enables more accurate identification of the non-zero channel-taps. In other words, for a 
$K$-sparse channel that needs to be estimated, leveraging the block structure allows for a wider range of $K$ while achieving similar estimation accuracy.
Meanwhile, BD-SI provides beneficial information for the identification of the non-zero channel-tap blocks, leading to a less restrictive upper bound on $K$ required for reliable recovery. Moreover, since more accurate identification leads to faster convergence, BD-CS exhibits a reduced computational complexity, even when BD-SI's computing resources are also taken into consideration.

\textcolor{black}
{Specifically, TABLE I presents the pros and cons of conventional CS, BCS and BD-CS enabled CE methods.}  The angular-domain and polar-domain BD-CS methods have performance advantages within different distance scenarios due to the various domain mappings. Note that the computational complexity of BD-SI cannot be simply measured because it is adjustable. As for the angular-domain BD-CS, it has the lowest complexity among all the methods compared, and may perform well when the distance is relatively high, i.e., being away from the BS. \textcolor{black}{As for the polar-domain BD-CS, although it has higher complexity than the angular-domain BD-CS, the joint exploitation of block structure and BD-SI results in higher accuracy and lower complexity than the polar-domain CS and BCS. Note that the lower complexity of BD-CS over BCS is that the more accurate selection of non-zero channel-tap blocks will accelerate the convergence of iterative algorithms.}  Meanwhile, the polar-domain BD-CS achieves the best estimation performance, when the distance is relatively small, i.e., being close to the BS. Thus, BD-CS enabled NF CE offers advantages in terms of both accuracy and complexity.

\section{BD-CS Enabled Communication Technology}\label{sec3}

Next, we provide the core meaning of PB, and put forth novel ideas for the NF CE and hybrid TPC solutions based on BD-CS.

\subsection{\textcolor{black}{Partition Boundary}} 

\textcolor{black}{As illustrated in Sec. \ref{II-C}, angular-domain BD-CS works well in the scenarios where the distance ingredient is relatively large, i.e., being far away from BS, while polar-domain BD-CS provides assured performance when the distance ingredient is relatively small, i.e., being close to BS. A fundamental question is to measure how far it is from the BS, which will facilitate the effective use of BD-CS based on different sparse mappings. To this end, PB is proposed to partition the NF region into inner and outer regions, where inner NF channel is close to BS and exhibits excessively weak sparsity, and outer NF channel is far away from BS and exhibits weak sparsity. PB is a distance metric primarily determined by $K$ in a $K$-sparse NF channel.}

Explicitly, PB is related to the upper bound, denoted by $\overline{K}^*$, of sparsity required for reliable recovery of angular-domain BD-CS \cite{liyang2022tsp}. $\overline{K}^*$ has a reciprocal relationship with the coherence of the measurement matrix, where coherence measures the degree of interdependence among the columns of the measurement matrix and is generally defined as the maximum inner product between any two different columns of the measurement matrix. Given a NF channel with $K$-sparsity, if $K>\overline{K}^*$, then the channel is within the inner NF region; otherwise, it is within the outer NF region. As $K$ decreases with increasing distance, a mapping between distance and $K$ can be established. Consequently, PB corresponds to the distance associated with $\overline{K}^*$.

\subsection{Channel Estimation}

Ensuring accurate CE is crucial for achieving high communication performance in extremely large-scale antenna arrays. 
In CS based CE approaches, the problem is usually transformed into sparse recovery based on a codebook \cite{dai2023cm}. To alleviate the severe weak sparsity in the angular-domain NF channel, a PDT matrix is designed in \cite{dai2022tcom} by sampling the angle uniformly, while sampling the distance non-uniformly. However, the complexity of PDT is relatively high, and conventional CS does not exploit the block structure and BD-SI, hence causing a potential performance gain erosion.

\begin{figure*}
	\centering
	\includegraphics[scale=0.65]{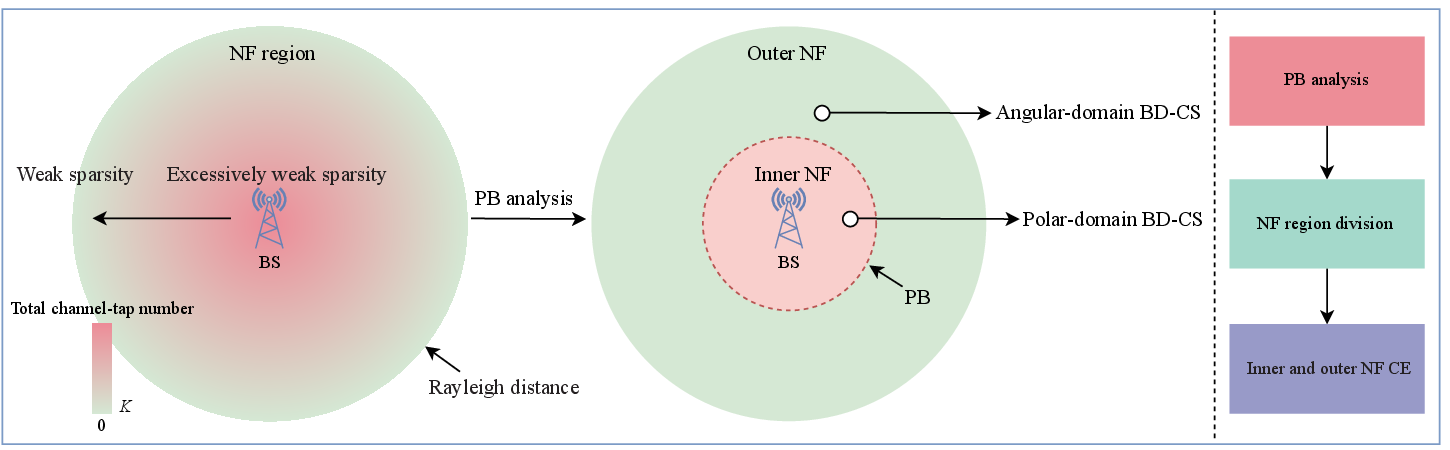}\\
	\caption{BD-CS aided CE.}\label{CEBD-CS}
\end{figure*}

The proposed CE by way of BD-CS is 
presented in Fig. \ref{CEBD-CS}. 

\begin{itemize}
\item The first step is to effectively partition the NF regions based on PB.

\item For an inner NF channel with $K$-sparsity, the weak sparsity is more severe and $K$ may be higher than the upper limit. BD-CS relies on the PDT for mapping a relatively sparse channel first, and then harnesses sparse CE. In this case, the low complexity of BD-CS can compensate for the high computing complexity of the PDT. For a $K$-sparse outer NF channel, CE can be carried out directly by BD-CS in the angular domain, since $K$ falls within the reliable recovery sparsity range of  BD-CS. Meanwhile, the angular-domain estimation avoids the weak sparsity caused by increasing the distance in PDT.
\end{itemize}

\subsection{Hybrid Transmit Precoding}

The design of hybrid TPC for conventional FF modeling is conveniently reformulated as a sparse reconstruction problem by exploiting the spatial structure of the channel. It allows the systems to approach their unconstrained performance limits, even when realistic transceiver hardware constraints are considered. 
For the NF scenarios, the authors of \cite{yzhang2022wcl} develop a two-phase beam selection method that sequentially searches for the optimal beam over the angle and distance ingredients. The study \cite{yjiang2023twc} exploits the beam scanning method and additionally proposes a maximum likelihood method and a focal scanning method for sensing the location of the receiver in the NF region. However, these studies do not consider the potential gain offered by the channel's block structure. Furthermore, they propose no prospective CS modeling based NF hybrid TPC for approaching the unconstrained performance limits.

\textcolor{black}{The proposed hybrid TPC by way of BD-CS is to obtain the optimal radio frequency and baseband precoding matrices $\mathbf{F}_{RF}$ and $\mathbf{F}_{BB}$ for maximizing the spectral efficiency, where the block sparsity of NF channels from different NF regions partitioned by PB is exploited.}
\begin{itemize}
	\item Based on the estimated NF channel, the first step is to harness channel decomposition using singular value decomposition, and obtain the optimal unconstrained unitary precoder $\mathbf{F}_{OPT}$ whose dimension is related to the number of valid data streams.
	
	\item For the inner NF, we formulate a cascade matrix, i.e., the measurement matrix, consisting of the continuous polar-domain steering vectors. Note that a certain number of columns in this measurement matrix could constitute the radio frequency TPC matrix $\mathbf{F}_{RF}$. The next objective is to accurately approximate the optimal unconstrained unitary TPC $\mathbf{F}_{OPT}$ based on the measurement matrix. To this end, BD-CS is utilized to obtain the sparse solution based on the measurement matrix and $\mathbf{F}_{OPT}$. The obtained sparse solution is the target baseband TPC matrix $\mathbf{F}_{BB}$, and the submatrix consisting of the steering vectors in the measurement matrix, which correspond to the non-zero coefficients in $\mathbf{F}_{BB}$, is the desirable $\mathbf{F}_{RF}$. For the outer NF, similar procedures can be employed but the measurement matrix is composed of the continuous angular-domain steering vectors.
	\end{itemize}

\subsection{Feasibility Verification}

\begin{figure}
	\centering
	\includegraphics[scale=0.4]{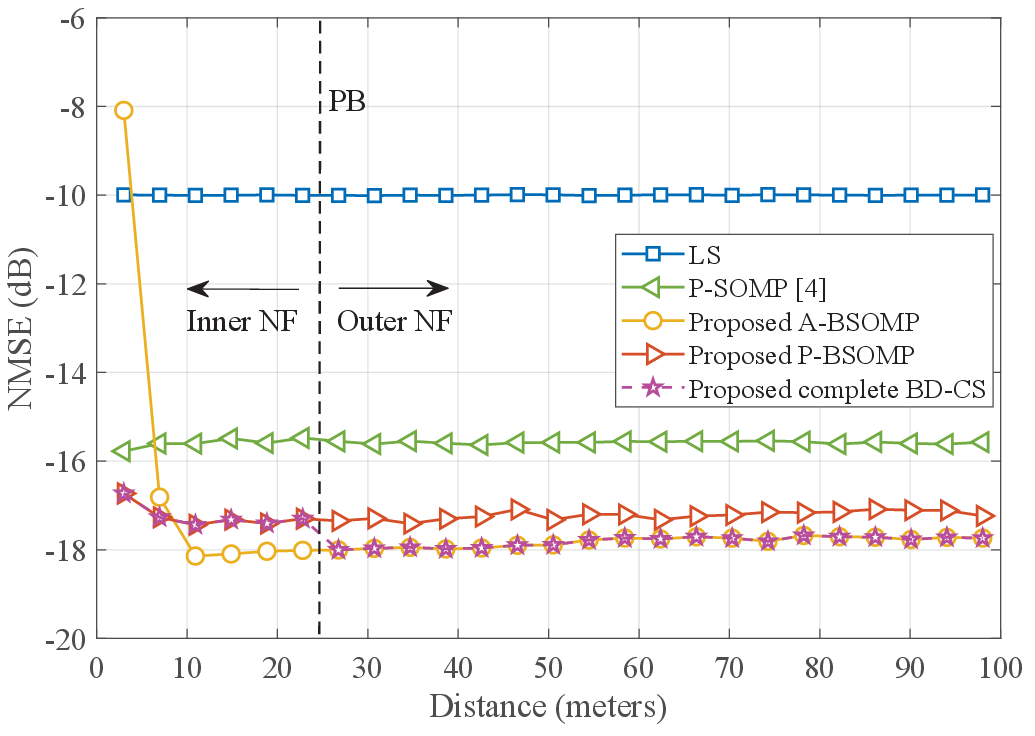}\\
	\caption{CE performance versus distance.}\label{simulation}
\end{figure}

\begin{figure}
	\centering
	\includegraphics[scale=0.31]{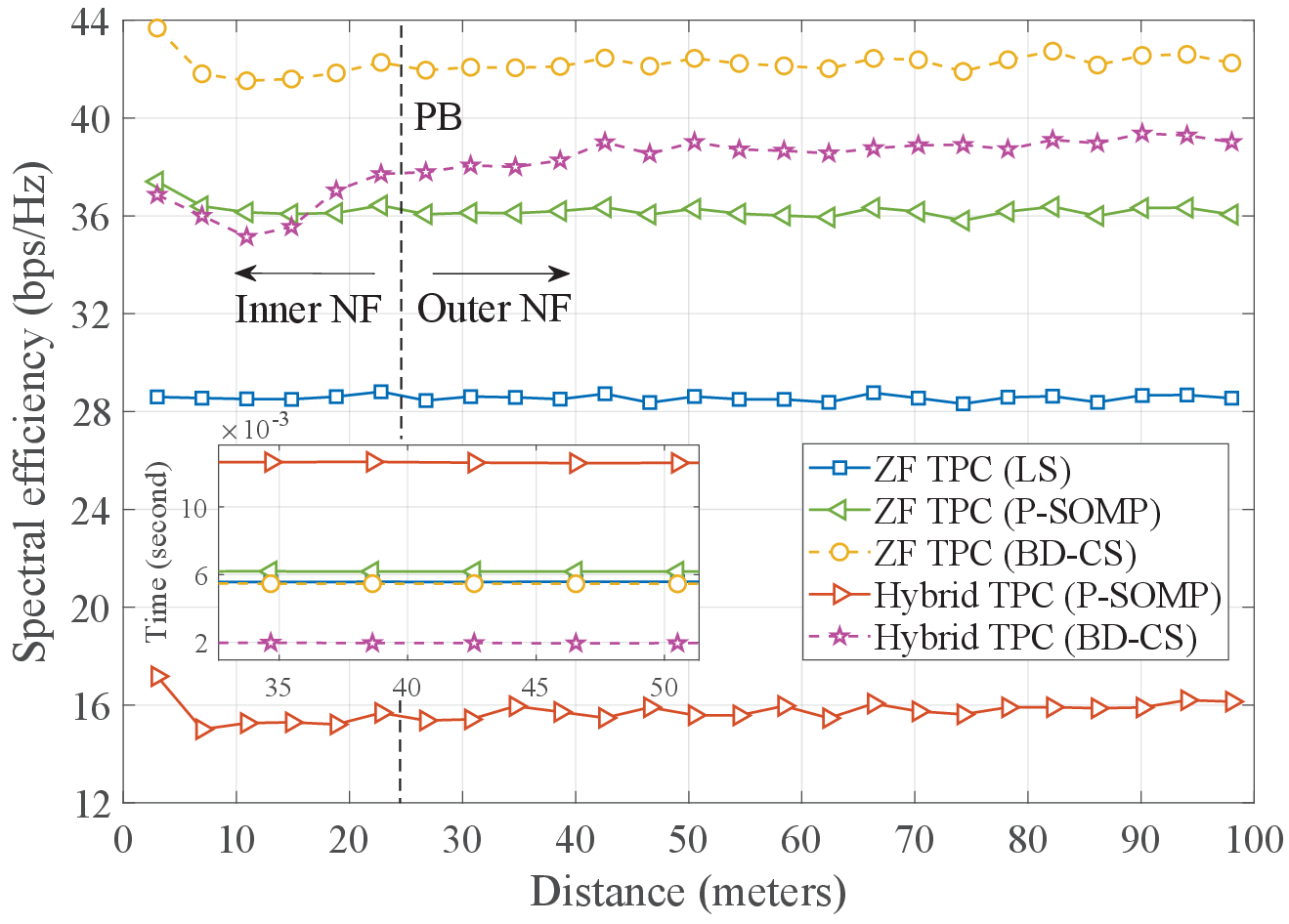}\\
	\caption{TPC performance versus distance.}\label{simulation2}
\end{figure}

This subsection provides an example to verify the feasibility of BD-CS assisted NF communications. A multi-user OFDM system is considered, where the number of the transmit antennas is $256$, the number of users is $8$, and the number of paths for each user is $6$. 
The BD-SI used in the simulations is the joint correlation property, and the number of subcarriers is $4$.  The results are presented in Figs. \ref{simulation} and \ref{simulation2}, which illustrates the performance of the normalized mean square error (NMSE) and the average sum rate as a function of the distance ingredient. 
In Fig. \ref{simulation}, we compare the NF CE performance regarding off-the-shelf algorithms, i.e., the classical least squares (LS) algorithm and the SOMP in polar domain (P-SOMP) \cite{dai2022tcom}, with the block SOMP in the angular domain (A-BSOMP), the block SOMP in the polar domain (P-BSOMP), and the complete BD-CS method. Note that conventional methods are independent of PB, while the BD-CS method represents a combination of results from A-BSOMP in the outer NF region and P-BSOMP in the inner NF region, as determined by PB analysis. The upper bound for sparsity required for reliable recovery is approximately 12 \cite{liyang2023arxiv}. Consequently, the PB can be calculated to be around 24 meters, based on the monotonicity of sparsity in DFT-based NF channels and the distance between users and the BS. It is evident that the complete BD-CS outperforms LS and P-SOMP algorithms in both the inner and outer NF regions. 

\textcolor{black}{In Fig. \ref{simulation2}, different TPC schemes are evaluated, containing
	zero-forcing (ZF) schemes using the estimated NF channels based on LS, P-SOMP and BD-CS, and the proposed hybrid TPC schemes using P-SOMP and BD-CS with the corresponding estimated channels. To better demonstrate the algorithm's ability to resist inter-user interference, the severe NF inter-user interference scenarios are considered, where NF channels of various users are similar. 
	The BD-CS assisted methods perform outer NF TPC using measurement matrices made up of angular-domain steering vectors, while they execute inner NF TPC with measurement matrices comprised of polar-domain steering vectors. 
	It can be observed that BD-CS assisted hybrid TPC scheme enjoys a stunning advantage of complexity while its TPC performance is close to that of BD-CS assisted ZF TPC which utilizes the fully connected architecture. In general, the TPC schemes by the way of BD-CS obtain more desirable spectral efficiency than that of P-SOMP assisted ZF TPC. Furthermore, LS based ZF TPC performs the worst among the ZF TPC schemes compared, even worse than hybrid TPC based on BD-CS, due to the absence of CS technique. When compared to P-SOMP assisted hybrid TPC that do not utilize the block sparsity of the NF channel, the results reveal that there is a non-negligible improvement  of the BD-CS enabled hybrid TPC in terms of both TPC performance and complexity, indicating that  BD-CS may be necessary for the efficient polar-domain based hybrid TPC.
	In a nutshell, combining the results of Figs. \ref{simulation} and \ref{simulation2}, BD-CS can provide efficient NF transmission by improving both CE and TPC performance.}

\section{Challenges and Opportunities}
\label{Challenges and Opportunities}

We present several future research directions for BD-CS enabled NF communications.

\subsection{NF Communication Theory with BD-CS}

The theoretical problems to be solved should involve the following three aspects. The first one is to study the less restrictive upper bound of sparsity required for reliable recovery of BD-CS for improved NF region partitioning. It is worth mentioning that the BD-SI aided theoretical analysis of the upper limit lacks systematic study, and the closed-form upper limit related to the distance ingredient is yet to be derived. Secondly, the necessary pilot overhead bounds of BD-CS have to be further developed, which may result in a low computational complexity. Note that the necessary pilot overhead bounds can be developed based on the sparsity limit analysis, which is regarded as a scalable theoretical extension. The third point is the analysis of spectral efficiency improvement. In fact, the BD-CS modeling is applicable to the multi-UE scenario. Its  channel capacity may be improved by utilizing multiple access technology, e.g., space-division multiple access \cite{dai2023cm}. Moreover, the exploitation of block structure in BD-CS enables more compact channel support information, which has the  potential to further improve the channel capacity.

\subsection{Reliability for Varying Clusters}

In practice, for a $K$-sparse channel, the value of $K$ and block structure characteristic of scatterer clusters undergo drastic changes, resulting in fluctuating CE and hybrid TPC solutions. From the angle of clusters, the mobile scatterers may separate or gather to break up an existing cluster or facilitate one new cluster. In this case, the block structure of the channel signal or the codebook matrix varies into a distinct one in terms of $K$ and block structure. The recovery may suffer from severe performance loss due to the imprecise inputs of the structural BD-SI. Thus, adaptive methods concerning the communication reliability should be developed based on the BD-CS enabled NF communication theory.  

\subsection{Modeling for Relay-Aided Communications}

In reconfigurable intelligent surface-aided NF communications, the conventional cascaded channel that only involves the phase ingredient of FF scenarios should be further extended to the distance ingredient related one. Fortunately, the concept presented in this article may also be applied to reconfigurable intelligent surface-aided systems. Moreover, the active reconfigurable intelligent surfaces can play the role of a transmitter or that of a relay. Hence the in-depth theoretical analysis can also follow the basic ideas of Sec. \ref{Challenges and Opportunities}-A with reconfigurable intelligent surface-aided NF communication model. 

\subsection{High-Frequency NF Localization}
Due to the susceptibility of high-frequency signals to attenuation and multipath effects, the low reliability of high-frequency near-field localization poses a tricky challenge that must be tackled.
For NF multi-target localization, the mobile devices to be located may constitute clusters in the area of interest. Based on the CS model formulation of multi-target localization, the measurement signals could be significantly affected by the interference. Fortunately, the signal to be recovered exhibits block sparsity. Meanwhile, similar BD-SIs, such as the position from the previous time slot and the joint correlation stemming from different sensing nodes, occur naturally, which can further improve the high-frequency localization reliability. Thus, BD-CS is able to provide accurate and real-time NF localization of the mobile device cluster. 

\section{Conclusions}
In this article, we investigate BD-CS technique for NF communications. Firstly, the block-sparse channel model is reformulated and BD-SI is obtained from three perspectives of the NF channel. 
 Secondly, both CE and TPC schemes by way of BD-CS are developed, where the NF region is partitioned into either inner or outer regions based on an elaborate PB.
The proposed BD-CS has superior performance in NF scenarios, and several promising research directions have been outlined. 

\ifCLASSOPTIONcaptionsoff
  \newpage
\fi

%\bibliographystyle{ieeetran}
%\bibliography{SBOLS}

\vspace {0.5cm}

\noindent\textbf{Liyang Lu} [M] received his Ph.D. degree from Beijing University of Posts and Telecommunications in 2022, China. He is currently a Post Doctoral Fellow with Tsinghua University. His research includes compressed sensing, cognitive radios, and MIMO communications.

\vspace {0.5cm} 

\noindent\textbf{Ke Ma} received his B.S. and Ph.D. degrees from the Department of Electronic Engineering, Tsinghua
University. His current research interests
include mmWave communications and intelligent communications.

\vspace {0.5cm} 

\noindent\textbf{Yue Wang} [SM] received his Ph.D. degree from Beijing University of Posts and Telecommunications, China. He is currently an 
Assistant Professor with the Department of Computer Science, 
Georgia State University, Atlanta, GA, USA. His interests include 
the areas of signal processing, machine learning, and wireless 
communications, with research focuses on trustworthy AI, 
compressed sensing, massive MIMO, millimeter-wave communications, wideband spectrum sensing, cognitive radios, direction-of-arrival estimation, high-dimensional data analysis, and 
distributed optimization and learning. He serves as Associate 
Editor for IEEE Transactions on Signal Processing.

\vspace {0.5cm} 

\noindent\textbf{Zhaocheng Wang} [F] received his B.S., M.S., and Ph.D. degrees from
Tsinghua University in 1991, 1993, and 1996, respectively. From 1996 to
1997, he was a Post Doctoral Fellow with Nanyang Technological University,
Singapore. From 1997 to 2009, he was a Research Engineer/Senior Engineer
with OKI Techno Centre Pte. Ltd., Singapore. From 1999 to 2009, he was a
Senior Engineer/Principal Engineer with Sony Deutschland GmbH, Germany.
Since 2009, he has been a Professor with Department of Electronic Engineering, Tsinghua University. He was a recipient of IEEE Scott Helt Memorial
Award, IET Premium Award, IEEE ComSoc Asia-Pacific Outstanding Paper
Award and IEEE ComSoc Leonard G. Abraham Prize.

\end{document}